\documentclass[twocolumn]{elsart3}
 \usepackage{graphics}
 \usepackage{graphicx}
 \usepackage{subfigure}
\usepackage{amssymb}
\begin{document}
\begin{frontmatter}
%
%
%
\title{Optical Readout in a Multi-Module System Test for the ATLAS Pixel Detector}
%
%
\author{Tobias Flick},
\ead{flick@physik.uni-wuppertal.de}
\author{Karl-Heinz Becks},
\author{Peter Gerlach},
\author{Susanne Kersten},
\author{Peter M\"attig},
\author{Simon Nderitu Kirichu},
\author{Kendall Reeves},
\author{Jennifer Richter}, and
\author{Joachim Schultes}
\address{University of Wuppertal, Gau\ss str. 20, 42097 Wuppertal, Germany}
\begin{abstract}
The innermost part of the ATLAS experiment at the LHC, CERN, 
will be a pixel detector, which is presently under construction. 
The command messages and the readout data of the detector are transmitted 
over an optical data path. 
The readout chain consists of many components which are 
produced at several locations around the world, and must work 
together in the pixel detector. To verify that these 
parts are working together as expected a system test has
been built up. It consists of detector modules,
optoboards, optical fibres, Back of Crate cards, Readout
Drivers, and control computers.\\
In this paper the system test setup and the operation of the readout 
chain is described. Also, some results of tests using the final pixel 
detector readout chain are given.
\end{abstract}
\begin{keyword}
ATLAS \sep Pixel \sep System Test \sep Multi-Module \sep Optical Readout \sep Optolink
\end{keyword}
\end{frontmatter}
%
\section{The ATLAS Pixel Detector}
\label{sec:pixeldetector}
The ATLAS pixel detector is the innermost detector of the ATLAS experiment \cite{TDR}. 
It is comprised of 3 barrel layers and 6 disks, with 3 disks on each side of the primary interaction point.
The three barrel shells are referred as B-Layer, Layer 1, Layer 2, from inside to outside. The shells are equipped with staves, each of which hold 13 modules glued on a carbon-carbon structure.  
The disks are built up of sectors which hold 6 modules on a carbon-carbon structure. Each disk has 8 sectors.\par
In total there will be 1744 modules having 46080 channels each. Every module is connected by a cable and a patch panel to an optoboard, which is then connected optically to the readout electronics.
\section{The Readout Chain}
\label{sec:readoutchain}
\begin{figure*}
   \centerline{
   \includegraphics[width=\textwidth, height=7cm]{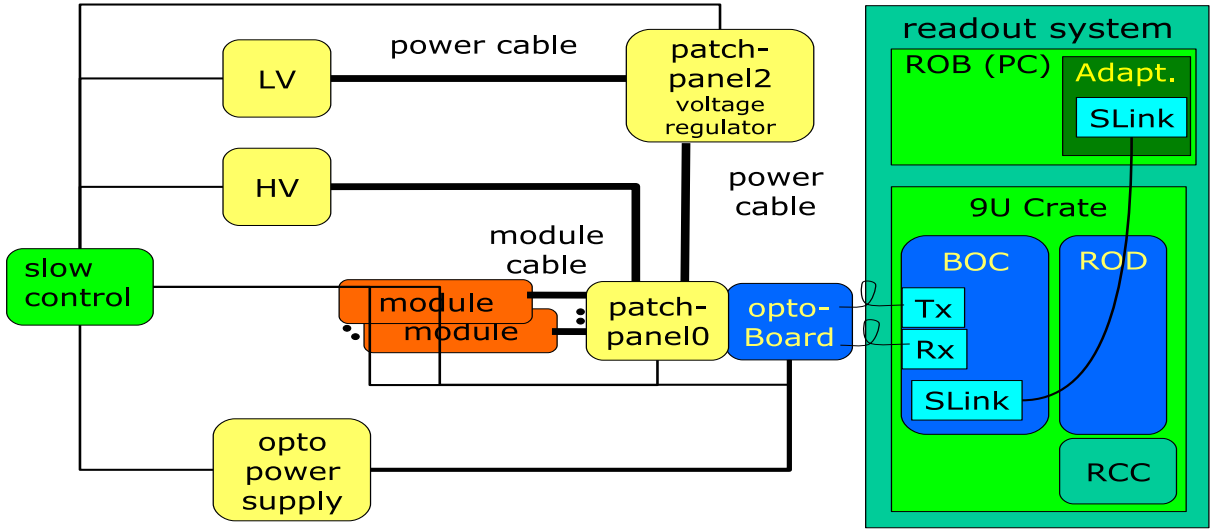}
   }
   \caption{Schematic of the system test setup (\cite{jens-systest})}
   \label{fig:setup}
\end{figure*}
The readout chain for the ATLAS pixel detector consists of an on-detector part and an off-detector part.
The off-detector components, located in the counting room, are the {\bf B}ack {\bf o}f {\bf C}rate cards (BOC), the {\bf R}ead{\bf o}ut {\bf D}rivers (ROD), the {\bf T}iming and {\bf I}nterface {\bf M}odules (TIM), and the {\bf R}ead{\bf o}ut {\bf B}uffers (ROB). The TIM receives the ATLAS clock and distributes it to the detector parts. The RoBs are storing the data. These last two devices are not part of the system test.\par
The on-detector components are the modules and the optoboards. The modules are connected by aluminium cables to  patch panels on which the optoboards are placed. Either 6 or 7 modules share one optoboard. The optoboards are connected with optical fibres to the off-detector electronics. A scheme of the system test setup is given in Figure \ref{fig:setup}.
\subsection{Readout Driver (ROD)}
In the counting room there are RODs\footnote{The Readout Driver has been developed by the Lawrence Berkeley National Lab and the University of Wisconsin, Madison, USA.} placed in 9U VME crates. The ROD is foreseen to perform the data formatting and the building of event fragments.
Additionally, it has capabilities to monitor the data taken. The card is controlled by a {\bf S}ingle {\bf B}oard {\bf C}omputer (SBC) placed in the same crate and acting as a {\bf R}eadout {\bf C}rate {\bf C}ontroller (RCC). For further information, please refer to \cite{ROD}.
\subsection{Back of Crate Card (BOC card)}
The Back of Crate card\footnote{The Back of Crate card has been developed by Cavensich Laboratory, Cambridge, UK. Wuppertal University, Germany, has  adopted it to the pixel system and is organising the production.} is placed back-to-back with the ROD. It is also a 9U VME-card and serves as the optical interface between the ROD and the optoboards. It has functionalities for data recovery, stream demultiplexing, and timing adjustment. Four transmission (TX) plug-ins and four receiver (RX) plug-ins can be mounted on the BOC card.
\begin{figure}[h] 
   \centering
   \includegraphics[angle=90,width=\columnwidth]{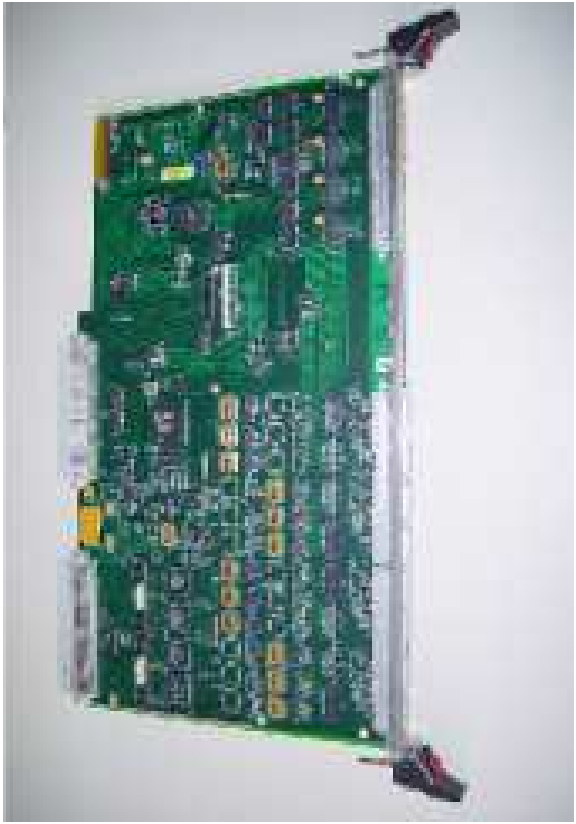}
   \caption{Back of Crate Card, the optical interface between Readout Driver and the modules}
   \label{fig:boc}
\end{figure}
\subsection{Opto plug-ins}
\begin{figure}[h]
   \centering
   \includegraphics[width=\columnwidth, height=4.5cm]{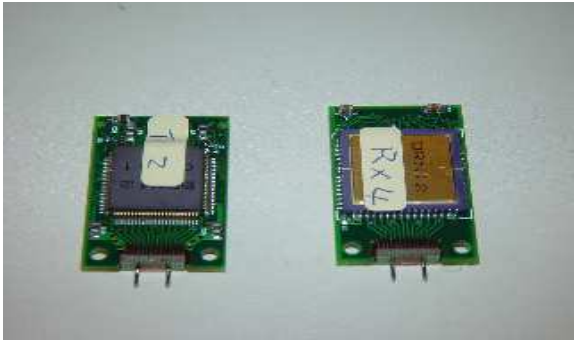}
   \caption{TX-plugin (left) housing the VCSEL array and RX-plugin receiving the optical signal from the optoboard. Plug-in dimensions: $2 x 3.5\,cm$}
   \label{fig:boc_80Mbit}
\end{figure}
On the pixel BOC card, there will be opto plug-ins\footnote{The plug-ins have been developed by Academica  Sinica, Taiwan.} to send/receive the data to/from the modules.
The transmitting device is the TX plug-in. It contains a {\bf B}i-{\bf P}hase-{\bf M}ark (BPM) chip to decode the clock and the commands for the modules to one single stream per module, and sends it optically via {\bf V}ertical {\bf C}avity {\bf S}urface {\bf E}mitting {\bf L}asers (VCSEL) to the detector.\par
The data from the modules are received by RX plug-ins. They contain a PiN\footnote{PiN: The name is derived from the structure as there is a {\bf P}-doped layer, an {\bf i}ntrinsic conducting part, and an {\bf N}-doped layer.}-diode and an amplifier chip, as well as the {\bf D}igital {\bf R}eceiver IC (DRX).\par 
Each of these devices has 8 channels, of which 6 or 7 will be used (see also \cite{txrx}). 
\subsection{Optoboard and Optical Fibres}
Two types of fibres will be used for the ATLAS pixel detector. There are radiation hard SIMM-fibres in the inner region of the ATLAS detector connected to less radiation tolerant GRIN-fibres in the outer region. The fibres will be installed as cables containing 8 ribbons with 8 fibres per ribbon. The distance between the optoboards and the opto plug-ins will be $\sim80\,m$.\par
\begin{figure}[h]
   \centering
   \includegraphics[width=\columnwidth]{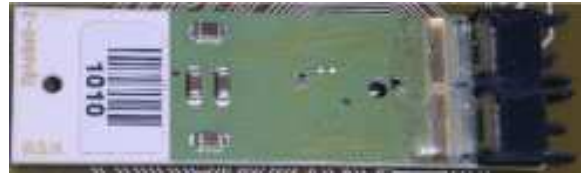}
   \caption{Optoboard, on the top side there are two VCSEL arrays with housings and driver chips (right side) and passive components. The optoboard measures $2\,cm\,x\,6\,cm$}
   \label{fig:oboard}
\end{figure}
The optoboard\footnote{The optoboard has been developed and produced by Ohio State University, USA and Siegen University, Germany.} (see Figure \ref{fig:oboard}) is the electrical-optical converter on the on-detector side. It receives the optical BPM-signal for the modules and converts it to two electrical signals, these being the clock and data lines.\\
In the opposite direction, the optoboard converts the electrical data signals from the modules into optical signals and then sends them to the RX plug-in mounted on the BOC card. Each optoboard can handle 7 modules. A detailed description is given in \cite{optoboard}.
\section{Bit Error Rate Measurement}
\label{bert}
 \begin{figure}[h]
    \centering
    \includegraphics[width=\columnwidth]{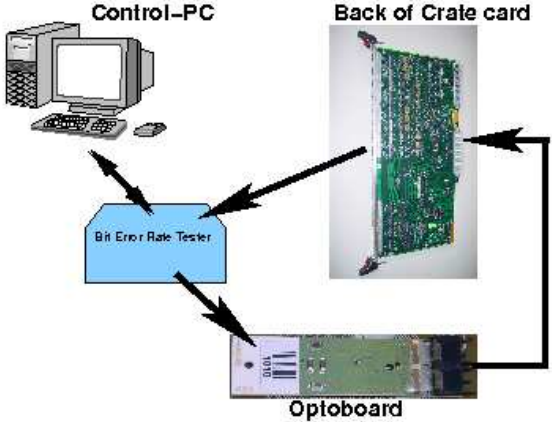}
    \caption{Setup for the bit error rate test}
    \label{fig:bertsetup}
 \end{figure}
\begin{table}
\centerline{
\begin{tabular}{ccccc}
\hline
\bf{Bandwidth} &  \bf{Number of } & \bf{Number of }    & \bf{Bit Error Rate}\cr
               &  \bf{Error Counts}          & \bf{Bits sent}              & \bf{Limit}\cr
\hline
$40\,Mb/s$     &  0                    & $15.832 \cdot 10^{12}$ & $6.32 \cdot 10^{-15}$\\ 
$80\,Mb/s$     &  0                    & $6.353 \cdot 10^{13}$  & $1.574 \cdot 10^{-14}$\\ 
\hline \\
\end{tabular} }
\caption{Results of the bit error rate measurements}
\label{tab:bert}
\end{table}
To test the quality of the optical data transmission system, a bit error rate  measurement was performed.
The link from the optoboard to the Back of Crate card was used to test the transmission. 
A test pattern from a Bit Error Rate Tester has been given to the optoboard electrically. This has then been transmitted over the optical link. The pattern received by the BOC card has been compared with the original one.\par
Two different bandwidths have been studied: 40\,Mb/s and 80\,Mb/s, as will be used in the ATLAS detector.
The test has shown a successful operation of the optical link with a good transmission quality. We measured no errors in the transmissions. Therefore only a limit for the bit error rate can be calculated. In the standard way one gets this out of equation \ref{eq:ber} assuming one error. The calculated limits for the bit error rate are $6.32\cdot10^{-15}$ for the 40\,Mb/s and $1.57\cdot10^{-14}$ for the 80\,Mb/s bandwidth. The results are listed in Table \ref{tab:bert}.
\begin{equation}
BER=\frac{\mbox{number of errors}}{\mbox{number of bits sent}}
\label{eq:ber}
\end{equation}
\section{System Test}
\label{systestsetup}
%
%
%
%
\begin{figure*}[t]
   \centerline{
	\subfigure[]{
		\includegraphics[width=\columnwidth, height=6cm]{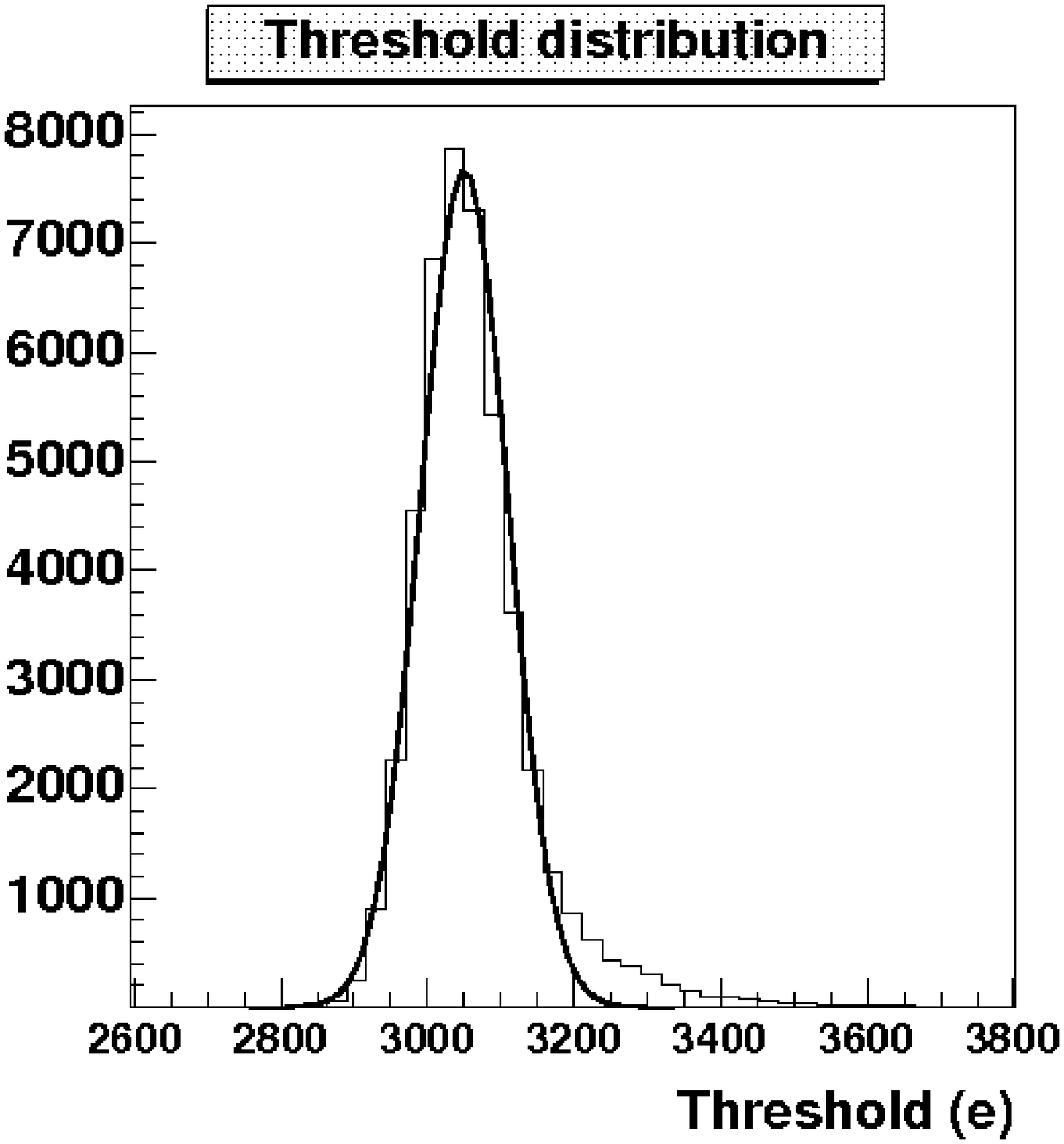}
   		\label{fig:tresh-histo}}
	\hfill
	\subfigure[]{
   		\includegraphics[width=\columnwidth, height=6cm]{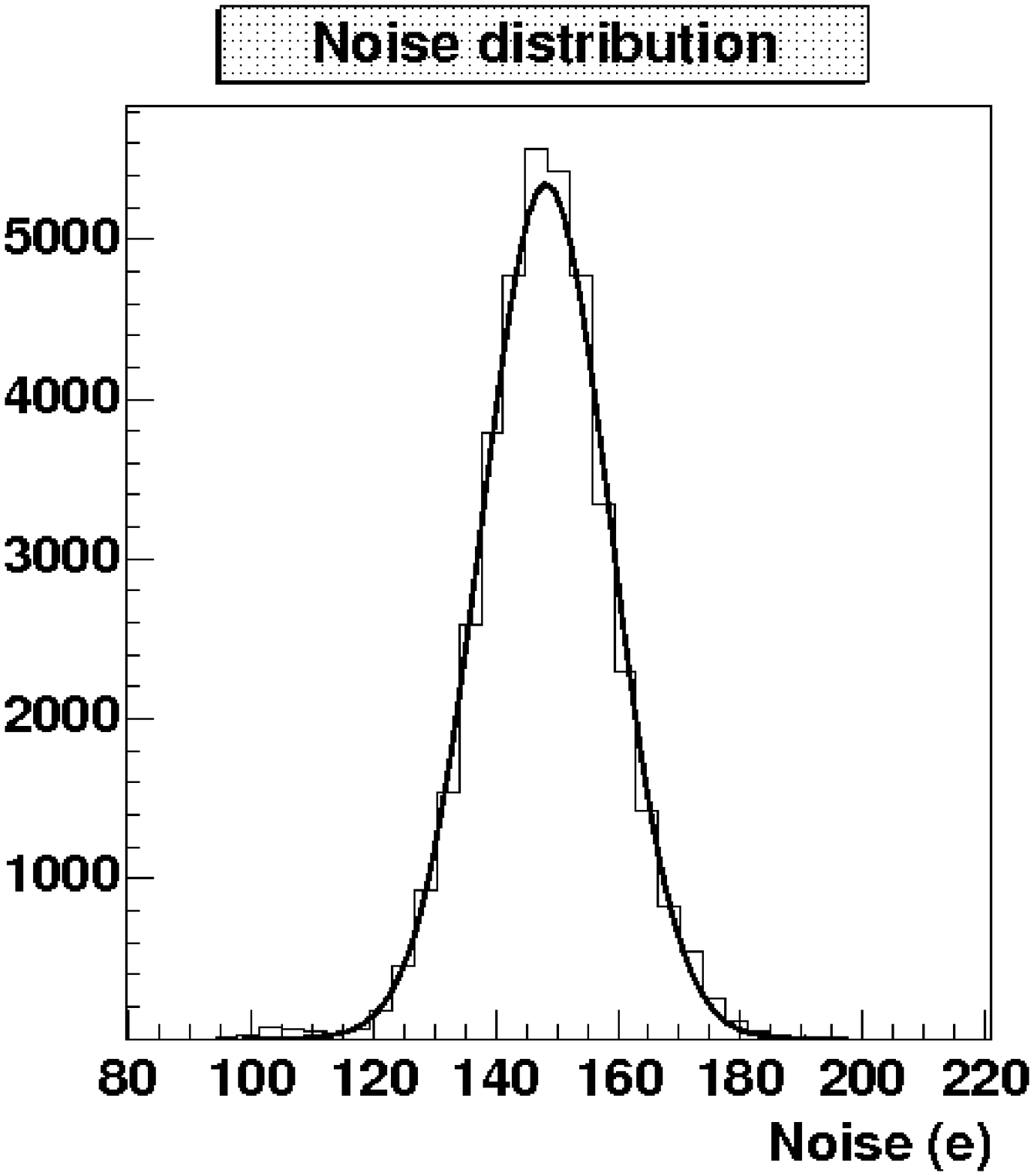}
		\label{fig:noise-histo}}}
\caption{(a) Threshold distribution histogram for a typical threshold scan. The mean threshold of 3051 electrons and the dispersion of 60 electrons are well in the specifications. (b) Noise distribution histogram for a typical noise scan. The mean noise of 150 electrons with a spread of 10 electrons is very good. The small dispersion indicates a uniform distribution over the whole module. }
\label{fig:thresh-scan}
\end{figure*}
%
%
\subsection{Setup}
The system test setup has been assembled to readout two staves, mounted together as a bi-stave. The bi-stave is mounted into a climate chamber to ensure operation under controlled temperature and humidity. The modules of each half stave are connected to a separate patch panel equipped with an optoboard. The power cables, which also connect to the patch panel, are very similar in terms of length and material to those which will be used in the experiment. This setup is used to study the behaviour of the modules on stave, the readout scheme, and the interaction of both. Of special interest for this paper is if there is any difference of the module behaviour due to the optical readout.\par
The powering as well as the monitoring of voltages, currents, and temperatures are performed by the {\bf D}etector {\bf C}ontrol {\bf S}ystem (DCS) (see also \cite{JoJo-DCS}). The system test DCS employs final prototypes. The interplay of the DCS and the readout system is studied as well.
\subsection{Test Results}
\label{testresults}
Several tests are necessary to check the module functionality. There are separate tests for the digital and for the analog part of the electronics.\par
From the analog scan (see Figures \ref{fig:tresh-histo} - \ref{fig:noise-histo}) one can derive the threshold and the noise of a module, which can then be compared to former measurements. More information about the module tests are given in \cite{jens-systest}.\par
\begin{figure}[b]
   \centering
   \includegraphics[width=\columnwidth]{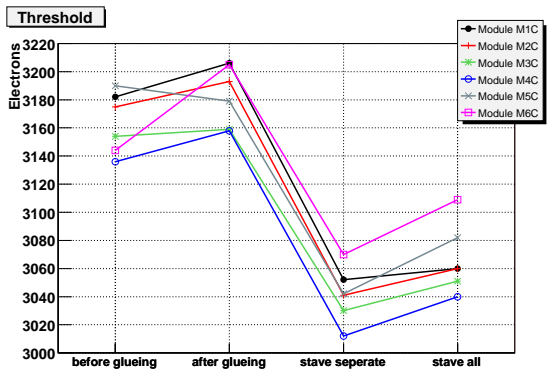}
   \caption{Comparison of the mean threshold for the different measurements}
   \label{fig:compthresh}
\end{figure}
\begin{figure}[b]
   \centering
   \includegraphics[width=\columnwidth]{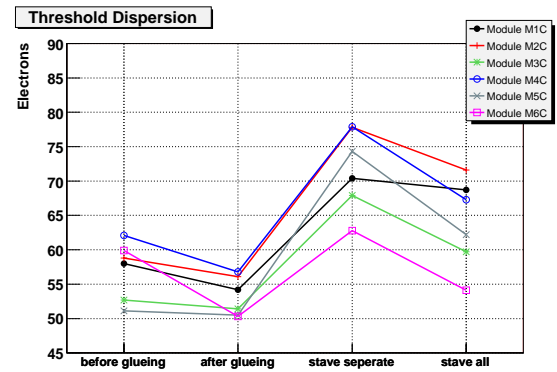}
   \caption{Comparison of the threshold dispersion for the different measurements}
   \label{fig:compdist}
\end{figure}
\begin{figure}[t]
   \centering
   \includegraphics[width=\columnwidth]{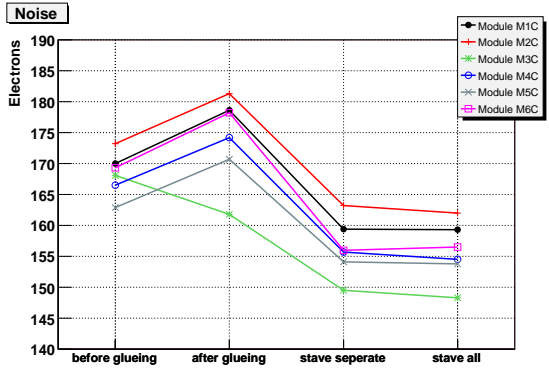}
   \caption{Comparison of the noise for the different measurements}
   \label{fig:compnoise}
\end{figure}

The measurements performed in the system test with one half of a stave (6 modules) have been compared to those performed for the modules individually during production:
\begin{itemize}
\item Using electrical readout only, the performance of the modules has been tested after module assembly (before glueing) and after full stave assembly (after glueing).
\item Using optical readout, tests have been done at stave level on a module-by-module mode (stave separate). The modules have been powered separately and they have been read out separately.
\item Using optical readout, tests with all modules together have been done at stave level as well (stave all). All the modules have been powered and were read out in parallel.
\end{itemize}
The results are shown in the plots given in Figures \ref{fig:compthresh}, \ref{fig:compdist}, and \ref{fig:compnoise}.
Small differences have been observed which can be attributed to temperature dependencies of the measurements. The electrical readout has been tested at $25^\circ\,C$ and $27^\circ\,C$ while the optical readout was performed at $18^\circ\,C$ and $19^\circ\,C$.\par
The threshold is tuned for a certain temperature. It rises with the operating temperature of the module. The dispersion of the threshold increases with the difference between tuning temperature and operating temperature. Finally, a lower operation temperature decreases the noise of the module.
\section{Summary}
\label{summary}
We built up a system test in Wuppertal to study the behaviour of a multi-module system in terms of powering, module behaviour, and readout. Special attention has been given to the optical data transmission on the readout chain and its influence on the module behaviour.\par
The results of the tests performed with the system test setup, including the optical readout, are in good agreement with those of previous ones. The modules behave as expected. One observes lower thresholds at lower temperatures. This effect is seen in the separate and common powering schemes. The dispersion is also changing with the change of thresholds. This one can be seen from Figure \ref{fig:compdist}.\par
The noise of the modules increases with the operation temperature. This effect can be seen in the different measurements. While the noise is around 170 electrons for the warmer environment ($\sim 25^\circ C$) it is around 160 electrons for the colder one ($\sim 19^\circ C$).\par
The behavior of the complete system has been studied. The transmission quality has been found to be good. The readout of a single module or of multiple modules is working as expected with the optical transmission system. There is no indication of degradation.\par
After showing that the small system is working stably, the system test will be scaled to a larger readout modularity.
\section{Acknowledgements}
The studies for the system test are a result of sharing knowledge and development work in the ATLAS pixel DAQ group. The software used is based on software packages which have been developed by several groups in Germany, Italy, Switzerland, and the USA. We would like to thank all people involved in this work.
%
%

%
\end{document}